\documentclass[3p,times]{elsarticle}
\usepackage{natbib}
\usepackage{graphicx}
\usepackage{epsfig,subfigure}
\usepackage{amssymb}
\usepackage{amsthm}
\usepackage{amsmath}

\journal{New Astronomy}
\begin{document}
\begin{frontmatter}
\title{WZ Sge: an eclipsing cataclysmic variable evolving towards the period minimum}

\author[1,2,3]{Han Z.-T.\corref{cor1}}
\ead {zhongtaohan@ynao.ac.cn}
\author[1,2,3]{Qian S.-B.}
\author[4]{Irina Voloshina}
\author[1,2,3]{Zhu L.-Y.}

\address[1]{Yunnan Observatories, Chinese Academy of Sciences (CAS), P. O. Box 110, 650216 Kunming, China.}
\address[2]{Key Laboratory of the Structure and Evolution of Celestial Objects, Chinese Academy of Sciences, P. O. Box 110, 650216 Kunming, China.}
\address[3]{University of Chinese Academy of Sciences, Yuquan Road 19\#, Sijingshang Block, 100049 Beijing, China.}
\address[4]{Sternberg Astronomical Institute, Moscow State University, Universitetskij prospect 13, Moscow 119992, Russia.}


\begin{abstract}

We present the photometric results of the eclipsing cataclysmic variable (CV) WZ Sge near the period minimum ($P_{min}$).
Eight new mid-eclipse times were determined and the orbital ephemeris was updated.
Our result shows that
the orbital period of WZ Sge is decreasing at a rate of $\dot{P}=-2.72(\pm0.23)\times{10^{-13}}\,s s^{-1}$. This secular decrease, coupled with previous detection of its donor,
suggest that WZ Sge is a pre-bounce system. Further analysis indicates that the observed period decrease rate is about $1.53$ times higher than pure gravitational radiation (GR) driving.
We constructed the evolutionary track of WZ Sge, which predicts that $P_{min}$ of WZ Sge is $\sim77.98 (\pm0.90)$ min. If the orbital period decreases at the current rate, WZ Sge will evolve past its $P_{min}$ after $\sim25.3$ Myr. Based on the period evolution equation we find $\dot{M}_{2}\simeq4.04(\pm0.10)\times10^{-11}M_{\odot}yr^{-1}$, which is compatible with the current concept of CV evolution at ultrashort orbital periods.

\end{abstract}

\begin{keyword}
binaries : eclipsing --
          binaries : evolution --
          stars : cataclysmic variables --
          stars : period bouncer--
          stars: individual (WZ Sge).
\end{keyword}
\end{frontmatter}

\section{Introduction}
The evolutionary theory of cataclysmic variables (CVs) predicts that there is an orbital period minimum ($P_{min}$) present and should be an accumulation of systems (i.e. "period spike") at $P_{min}$ (Paczy{\'n}ski \& Sienkiewicz 1981; Kolb \& Baraffe 1999). The relatively stable mass transfer in CVs is driven by angular momentum losses (AMLs). As a CV evolves, the system requires to shrink its orbit in order to keep the Roche lobe in touch with the donor, resulting in the orbital period decrease. When the system's period reduces to $P_{min}$, the secondary star is driven out of thermal equilibrium. At this point, the donor becomes a substellar object. The transition from a low-mass star to substellar object leads to its radius to increase in response to mass loss. As a result, the binary separation must expand in order to adapt the internal structure change of the donor. After that, the systems pass through $P_{min}$ and evolve into so-called period bouncers. These predictions have been confirmed by G$\ddot{a}$nsicke (2009) using the Sloan Digital Sky Survey CVs, who located the position of "period spike" at $82.4(\pm0.7)$ min.

In the standard model, most of CV systems ($\sim70\%$) were thought to be period bouncers (Kolb 1993). However, only a few period bouncer candidates were reported until now by detecting the brown dwarf secondaries (e.g., Littlefair et al. 2006, 2008; Zharikov et al. 2008; Savoury et al. 2011; McAllister et al. 2015). As one of period bouncer candidates, WZ Sge has a short orbital period of $81.6$ min (Patterson et al. 1998) with a low-mass secondary star ($M_{2}<0.11M_{\odot}$) (Steeghs et al. 2001), which is close to the hydrogen-burning limit. More recently, its donor was estimated to be a L-dwarf by Harrison (2016), suggesting that WZ Sge may not be a period bouncer. To identify a period bouncer, the sub-stellar donor is just one of necessary conditions. Therefore, more evidence for identifying period bouncers was required. Fortunately, WZ Sge is an eclipsing CV with a high inclination of $\sim77^{\circ}$ (Steeghs et al. 2007). The eclipsing nature provides a rare opportunity to ascertain its evolutionary state. In eclipsing CVs, the most common method for studying the orbital period changes is the timing method by analyzing their observed minus calculated ($O-C$) diagrams ("O" refers to the observed actual mid-eclipse times, while "C" represents those calculated with a given linear ephemeris.). An $O-C$ diagram is the $O-C$ values plotted against the integer E, from the ephemeris. E is the number of orbital cycles that occurred between the two eclipse times.
In recent years this method has been used to test the evolutionary status of several short-period eclipsing CVs such as Z Cha (Dai et al. 2009), OY Car (Han et al. 2015), V2051 Oph (Qian et al. 2015). Previous studies revealed that no long-term change in $O-C$ curve of WZ Sge was found (Robinson et al. 1978; Skidmore et al. 1997; Patterson et al. 1998). In the present paper, we present new CCD photometric observations of WZ Sge and detect a secular decrease in the $O-C$ diagram. Then its evolutionary state was discussed and some parameters were constrained by comparing with theoretical models.

\section{Observations}

The observations were obtained by using the 85-cm reflecting telescope mounted an Anor DW436 1K CCD camera at the XingLong station of the National Astronomical Observatory  and with the 2.4-m telescope at the Lijiang observational station of Yunnan Observatories, from 2008 to 2016. During the observations, no filters were used in order to improve the time resolution. All observed CCD images were reduced by applying the aperture photometry package of IRAF. Differential photometry was performed, with a nearby non-variable comparison star. Two eclipsing profiles of WZ Sge are displayed in Fig. 1. To get more mid-eclipse times, WZ Sge was monitored with the Sino-Thai 70-cm reflecting telescope at the Lijiang observational station. This telescope is equipped with an Andor DW936N 2K CCD camera. A good eclipsing profile was obtained and is displayed in Fig. 2.
As shown in Fig. 1 (and Fig. 2), the out-of-eclipse shapes are variable with time. In addition to the profile changes, the light curves also show the rapid oscillations in brightness, which may be associated with accretion events. The mid-eclipse times are determined by using the same method as in Robinson et al. (1978). In this method, the mid-eclipse times are the mean of one-half flux times during eclipse ingress and egress. The errors are the standard errors in measuring mid-eclipse times, and they depend on the time resolution and signal-to-noise ratio during observations. All mid-eclipse times and their errors are listed in Table 1. The eclipse width was calculated as $184\pm6$ s, which is close to $164\pm9$ s estimated by Robinson et al. (1978) and $210\pm20$ s in Patterson et al. (1998).

\begin{figure}[!h]
\begin{center}
\includegraphics[width=1.0\columnwidth]{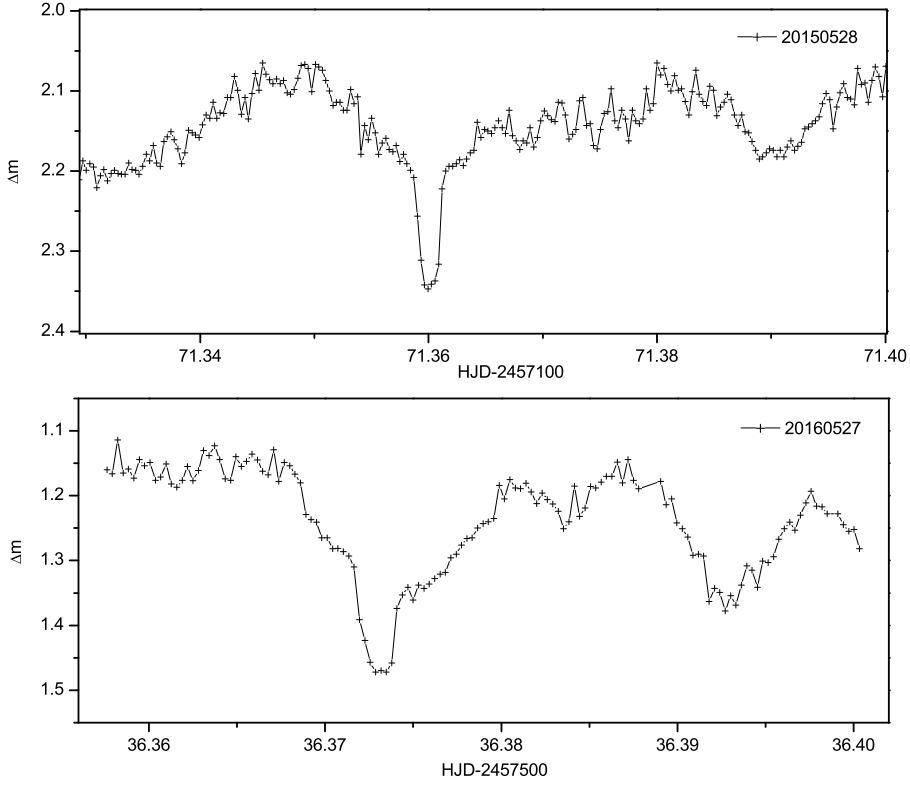}
\caption{Two eclipsing light curves of WZ Sge obtained with 2.4-m telescope in China.}
\end{center}
\end{figure}

\begin{figure}[!h]
\begin{center}
\includegraphics[width=1.0\columnwidth]{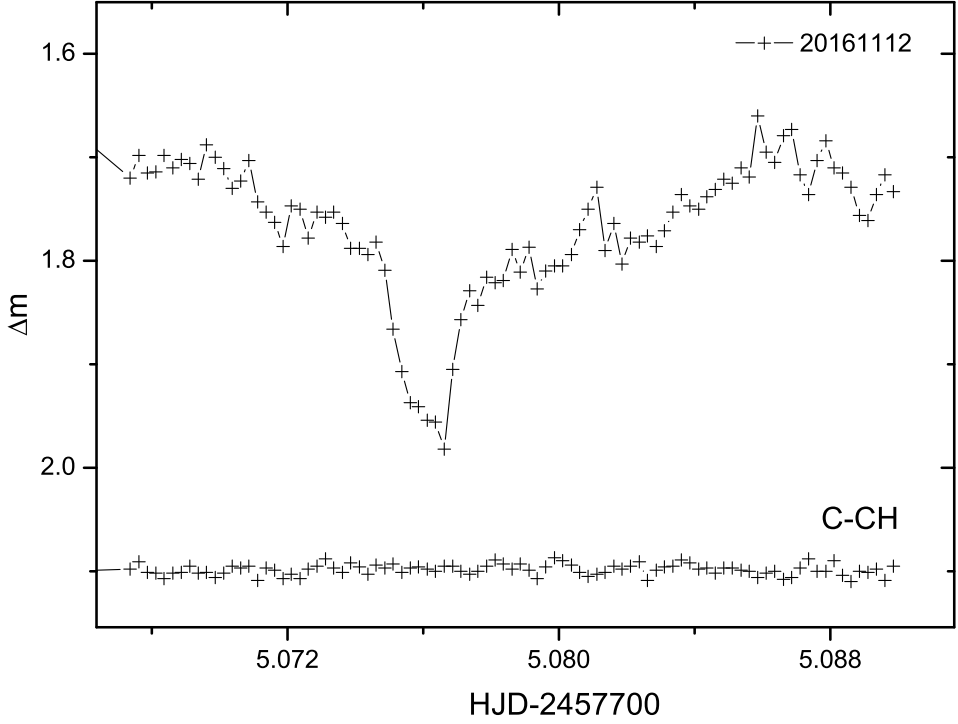}
\caption{An eclipsing profile of WZ Sge in the N band obtained with Sino-Thai 70-cm telescope on 12 November, 2016. The symbol "C-CH" refers to the magnitude differences between the comparison and the check stars.}
\end{center}
\end{figure}

\begin{table*}
\caption{New CCD mid-eclipse times of WZ Sge. }
 \begin{center}
 \small
 \begin{tabular}{ccccccc}\hline\hline
Date            &Min.(HJD)        &E          &O-C          &Errors        &Filters      &Telescopes      \\\hline
2008 Nov 07     &2454777.99890    &303950	  &-0.00039	   &0.00010     &N            &85cm            \\
2008 Nov 10     &2454781.00332    &304003	  &-0.00043	   &0.00010     &N            &85cm            \\
2008 Nov 26     &2454796.98958    &304285	  &-0.00015	   &0.00005     &N            &2.4m            \\
2008 Nov 29     &2454799.99398    &304338	  &-0.00020	   &0.00005     &N            &2.4m            \\
2008 Nov 30     &2454801.01440    &304356	  &-0.00016	   &0.00005     &N            &2.4m            \\
2015 May 28     &2457171.35995    &346170	  &-0.00021	   &0.00005     &N            &2.4m            \\
2016 May 27     &2457536.37287    &352609	  &-0.00034	   &0.00005     &N            &2.4m            \\
2016 Nov 12     &2457705.07587    &355585     &-0.00037    &0.00005     &N            &70cm            \\\hline
\end{tabular}
\end{center}
\end{table*}

\begin{figure}[!h]
\begin{center}
\includegraphics[width=1.0\columnwidth]{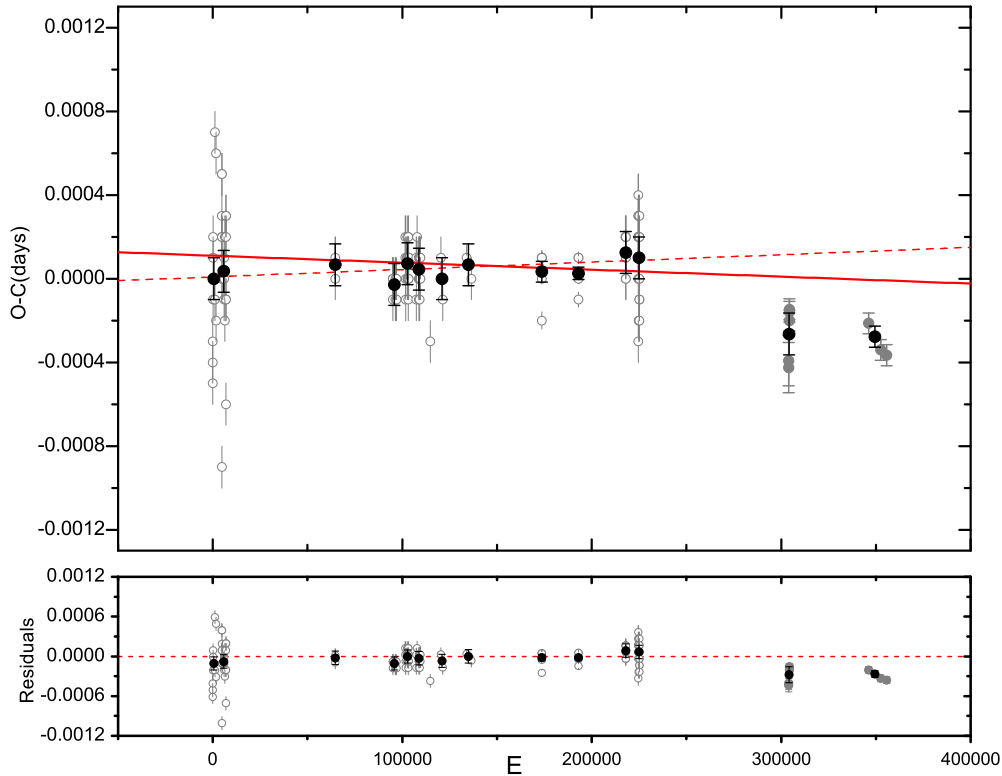}
\caption{$O-C$ diagram of WZ Sge constructed with a linear ephemeris. The grey open circles and solid dots refers to the data in literature and our observations, respectively.
The black solid dots are the mean values of each data set. The dashed red line in the upper panel is the linear fit derived by Patterson et al. (1998). The red solid line represents the linear ephemeris from our best fitting. The lower panel displays the fitting residuals from the complete linear ephemeris.}
\end{center}
\end{figure}

\begin{figure}[!h]
\begin{center}
\includegraphics[width=1.0\columnwidth]{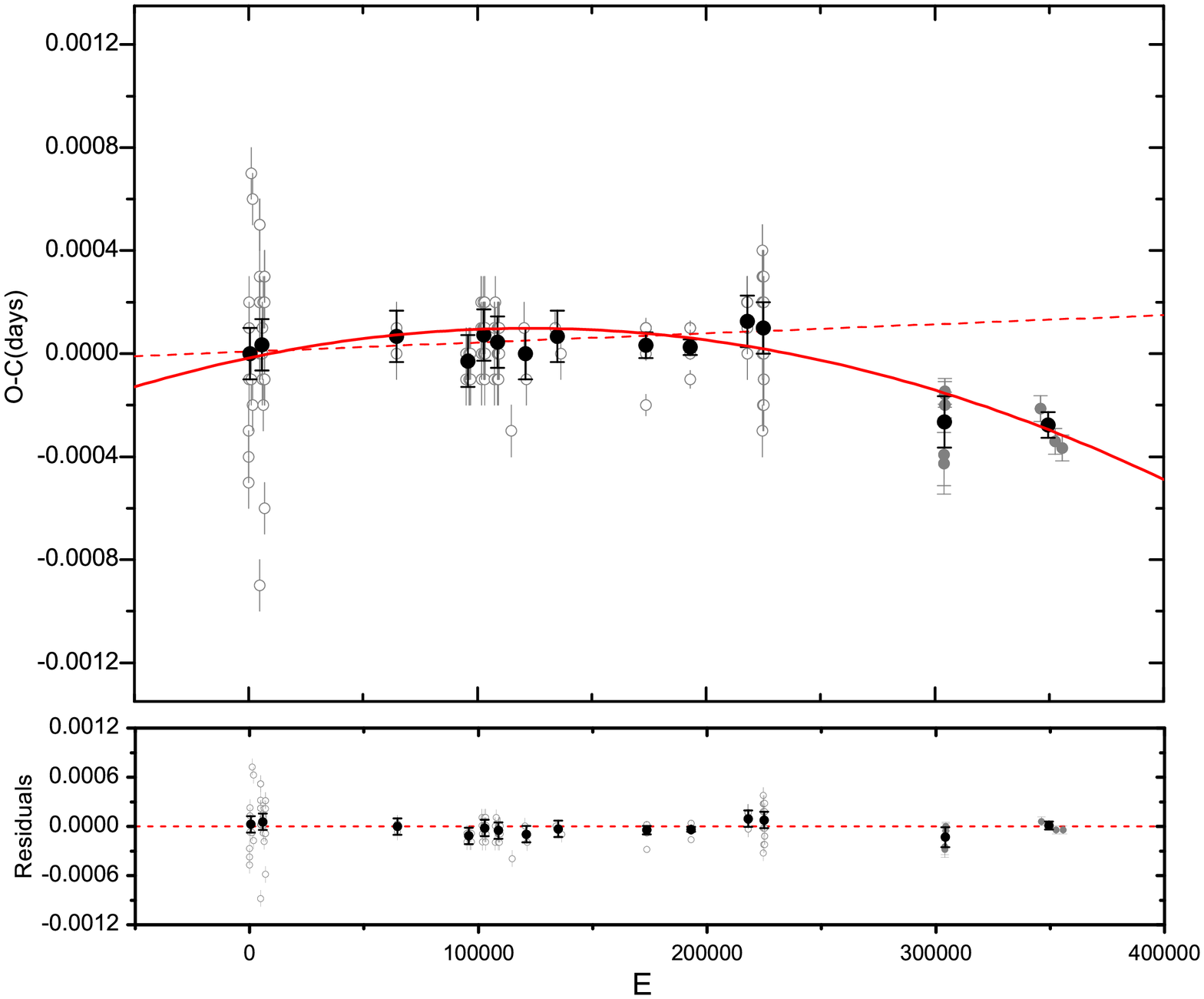}
\caption{$O-C$ diagram of WZ Sge. The grey open circles, grey solid dots with error bars and the dashed red line in the upper panel have the same meaning as in Fig. 2. The red solid line refers to the quadratic ephemeris from our best fitting. After the downward parabolic change was removed, the residuals are plotted in the lower panel.}
\end{center}
\end{figure}

\section{Results}

Mid-eclipse times of WZ Sge have been published in the literatures and the orbital period analyses have been presented by several authors. Robinson et al. (1978) found that no sign of any orbital period change. Later, Skidmore et al. (1997) updated the orbital ephemeris and suggested that the long-term evolution in the orbital period cannot be detected. After just one year, Patterson et al. (1998) revised the ephemeris again and indicated that the best description for $O-C$ is still a linear ephemeris.

Using our new data (see Table 1) together with all of timings in the literatures, the latest version of $O-C$ diagram is displayed in Fig. 3 (and Fig. 4). The $O-C$ values of all observed timings were computed with the linear ephemeris given by Patterson et al. (1998):
\begin{equation}
Min.I = HJD\,2437547.7284+0.056687846\times{E},
\end{equation}
where HJD\,2437547.7284 is the initial epoch and 0.056687846 d is orbital period.
Note that the updated $O-C$ diagram gives a baseline of $\sim55$ yrs.
We removed some mid-eclipse times with quoted errors larger than 0.001 days in our analysis.
To clearly show the change trend of the $O-C$ curve, the statistical average method was used to calculate the mean of all data segments (black solid dots in Fig. 3 and 4)
As shown in the top panel of Fig. 3, the new data (gray solid dots) don't follow the previously predicted constant-period ephemeris (dashed red line), implying that a simple linear ephemeris may not be a good description. Nevertheless, it cannot be apriori excluded. Thus a linear least-squares fit was first used to represent the $O-C$ curve. The solid red line in the upper panel of Fig. 3 refers to the best-fitting linear ephemeris for the latest $O-C$ diagram.
Clearly, the observed data reveal significant deviations from this ephemeris, especially for new data (gray solid dots), and the residuals show a possible secular decrease (see lower panel of Fig. 3). It seems that a quadratic ephemeris can describe the general trend of the $O-C$ well. We added a quadratic term to the ephemeris and found that it is a much better fit than the linear ephemeris (solid red line in Fig. 4). During the analysis, the weighted least-squares method was used and the weights for all mid-eclipse times were scaled as the inverse squares of their errors.
All results of the least-squares fitting are summarized in Table 2. In order to describe the goodness-of-fit of the two models, we calculated the R-square values, which are listed in Table 2. The R-square values also suggest that the quadratic ephemeris fits the data much better than the linear ephemeris.
Moreover, an analysis of variance (i.e. F-test) is used to check whether the quadratic fit represents a significant improvement over the linear fit (e.g. Pringle 1975). The F-test results are also listed in Table 2, which indicate that the quadratic term is significant, well above 99.99\% level. Thus, the best-fit of the O-C diagram for WZ Sge is a quadratic ephemeris.
Fitting residuals are plotted in the lower panel of Fig. 4. A downward parabola in upper panel of Fig. 4 indicates a secular decrease at a rate of $\dot{P}=-1.54(\pm0.13)\times{10^{-14}}\,days/cycle=-2.72(\pm0.23)\times{10^{-13}}\,s s^{-1}$.

\begin{table*}
\caption{Parameters of the best fitting for $O-C$.}\label{elements}
\begin{center}
\small
\begin{tabular}{llll}
\hline
\hline
                                                               &Linear ephemeris:                     &&Quadratic ephemeris:  \\
Parameters                                                     &$O-C=\Delta{T_{0}}+\Delta{P_{0}}{E}$  &&$O-C=\Delta{T_{0}}+\Delta{P_{0}}{E}+{\beta}{E^{2}}$\\
\hline
Degrees of freedom, $\nu$                                      & 148                                  &&147                                  \\
Correction on the initial epoch, ${\Delta{T_{0}}}$ (days)      & $1.09(\pm0.30)\times10^{-4}$         && $2.87(\pm0.38)\times10^{-5}$        \\
Correction on the initial period, ${\Delta{P_{0}}}$ (days)     & $-3.34(\pm1.70)\times10^{-10}$       && $+1.96(\pm0.43)\times10^{-9}$       \\
Rate of the linear decrease, $2\beta$ (days/cycle)             &                                      && $-1.54(\pm0.13)\times10^{-14}$  \\
R-square of Goodness of Fit                  &$0.025$              && $0.230$          \\
Adjust R-square                            &$0.019$                            && $0.219$          \\
F-statistic value, $F$                        &$3.83$               && $21.77$          \\
$P$ value of F-test, $P_{rob}>F$                &$0.052$              && $5.31\times10^{-9}$          \\
\hline
\end{tabular}
\end{center}
\end{table*}

\section{DISCUSSION}

When a short-period CV evolves into a period bouncer, its orbital period should increase and the donor becomes a sub-stellar object.
With an orbital period near the period minimum and a late-type donor,  WZ Sge was classified as a possible period bouncer by Patterson (1998).
Meanwhile, Ciardi et al. (1998) claimed that WZ Sge has passed $P_{min}$ and the donor is a sub-stellar object with a low temperature of $\leq1700 K$.
Until recently, a direct detection for the donor of WZ Sge led to the suspicion that it may not be a bounce-back system (Harrison 2016). Harrison pointed out that the L2-L5 donor is earlier than the predicted
spectral type in period bouncers by Knigge et al. (2011). However, this solution still has some deficiencies. First, the predicted result by Knigge et al. (2011) is based on a semi-empirical donor evolution sequence, and the samples contain the intrinsic dispersions. Second, the L2-L5 donor was detected by using $K$-band spectra. In effect, $J$-band is much better than $K$-band for L-dwarf identification. Therefore, further evidence should be provided to confirm WZ Sge's classification. As noted above, the orbital period of WZ Sge is decreasing at a rate of $\dot{P}=-2.72(\pm0.23)\times{10^{-13}}\,s s^{-1}$. If it is a period-bounce system, the period should be increasing rather than decreasing. Combining the decreasing period with a L2-L5 donor presented by Harrison (2016), we believe that WZ Sge is a pre-bounce CV and has not yet evolved past $P_{min}$.

WZ Sge-type CVs are generally thought to be the dominant objects at $P_{min}$ (Zharikov 2014). As a prototype of these stars, WZ Sge has a short period of $81.6$ min, which is close to the current estimated $P_{min}$ of $\sim81.8\pm0.9$ min by Knigge et al. (2011). Hence, its evolutionary state plays a crucial role for understanding the evolution of CVs and testing theoretical models. In the standard model of CVs, the secular evolution of short-period CVs ($P_{orb}\leq2$ h) is driven purely by gravitational radiation (GR) (Rappaport et al. 1983; Spruit \& Ritter 1983). The period decrease rate due to GR was given by (Kraft et al. 1962; Paczy{\'n}ski 1967):
\begin{equation}
\frac{\dot{P}_{GR}}{P}=-3\frac{32G^{3}}{5c^{5}}\frac{M_{1}M_{2}(M_{1}+M_{2})}{a^{4}}.
\end{equation}
$a$ is the orbital separation, $M_1$ and $M_2$ are the masses of gainer and donor stars, respectively.
For WZ Sge, $M_1=0.85(\pm0.04)M_{\odot}$ and $M_2=0.078(\pm0.006)M_{\odot}$ given by Steeghs et al. (2007), which can be combined with Kepler's third law to yield $a=0.61R_{\odot}$.
Finally, the period decrease caused entirely by GR is calculated as $\dot{P}_{GR}=-1.78(\pm0.12)\times{10^{-13}}\,s s^{-1}$.
The observed period decrease is $\simeq1.53(\pm0.11)$ times higher than the purely GR-driven decrease rate. To ascertain the evolutionary status of WZ Sge, we have studied the relation between donor mass and orbital period (see Fig. 5). The different evolutionary tracks are shown in Fig. 5. The blue track represents the revised model from Knigge et al. (2011) and the red dashed is the standard model track. The revised model showed that AML rates below the gap are $\simeq2.47$ times higher than pure GR driving. The location of WZ Sge ($\simeq1.53\times GR$) lies between the revised model and the standard model track. This position corresponds to $M_2=0.078M_{\odot}$, in quite good agreement with the result of Steeghs et al. (2007).
Moreover, we have constructed the possible evolution track of WZ Sge (green dash line) using the method described by Knigge et al. (2011). Along this track, the predicted $P_{min}$ for WZ Sge is $\sim77.98(\pm0.90)$ min. Assuming the orbital period decreases at the current rate ($\dot{P}$), then WZ Sge will evolve past its $P_{min}$ after $\sim25.3$ Myr.

\begin{figure}[!h]
\begin{center}
\includegraphics[width=1.0\columnwidth]{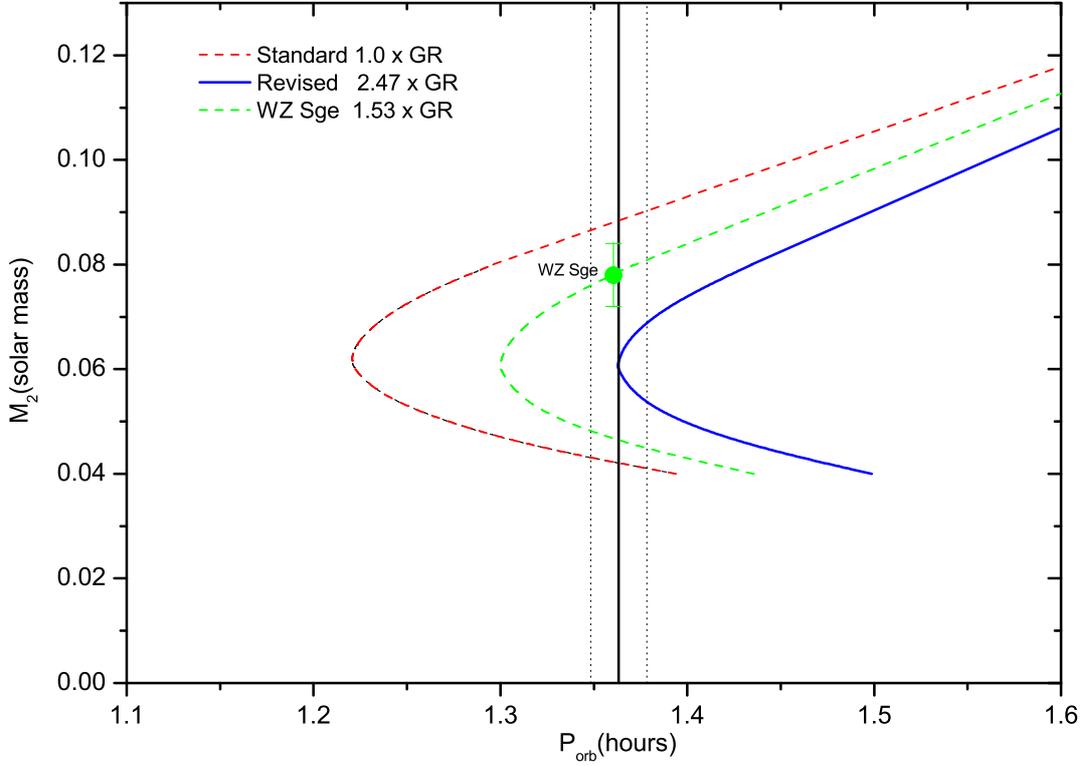}
\caption{Donor mass ($M_2$) versus orbital period $(P_{orb})$ relationship for WZ Sge. The location of WZ Sge is labelled as the solid green circle. The evolutionary track according to the standard model of CV evolution (pure GR driving AML below the gap) is shown with the dashed red line. The solid blue line represents the revised model track from Knigge et al. (2011). The vertical solid black line marks the location of the period minimum ($\sim81.8\pm0.9$ min) predicted by the revised model. The dashed green line is the possible evolutionary track of WZ Sge.}
\end{center}
\end{figure}

$P_{min}$ is closely connected to the nature of CV donors. Based on the mass-radius and period-density relation for CV donors, the period evolution equation was derived as (Robinson et al.1991; Knigge et al. 2011)
\begin{equation}
\frac{\dot{P}}{P}=\frac{3\zeta-1}{2}\frac{\dot{M_2}}{M_2},
\end{equation}
where $\zeta$ is the mass-radius index. For the orbital period decrease ($\dot{P}<0$), $\zeta>\frac{1}{3}$ is required. When $\zeta=\frac{1}{3}$, $P_{orb}=P_{min}$.
A revised model fit to the mass-radius relation of CV donors provides the relevant parameters including mass-radius index (Knigge et al. 2011). For WZ Sge, the donor mass corresponds to $\zeta\simeq0.559$. Plugging these parameters into equation (3), we find $\dot{M}_{2}\simeq-4.04(\pm0.10)\times10^{-11}M_{\odot}yr^{-1}$.
The position of WZ Sge-type stars is marked on $\dot{M}_{2}$ versus $P_{orb}$ diagram of CVs by Zharikov et al. (2013), corresponding to the mass transfer range from $\sim2\times10^{-11}M_{\odot}yr^{-1}$
to $\sim8\times10^{-11}M_{\odot}yr^{-1}$. Our $\dot{M}_{2}$ value is compatible with this limit. The decreasing of orbital period of WZ Sge is accompanied by the decline in the mass transfer rate. After reaching a lower mass transfer rete ($\dot{M}_{2}\leq2\times10^{-11}M_{\odot}yr^{-1}$), it eventually enters the period bouncer regime.

\section{CONCLUSIONS}

We have presented the photometric results of the eclipsing CV WZ Sge. Seven new mid-eclipse times were determined and
the updated ephemeris is significantly improved and allows to ascertain the evolutionary state of WZ Sge. Our analysis shows that
the orbital period of WZ Sge is undergoing a secular decrease at a rate of $\dot{P}=-2.72(\pm0.23)\times{10^{-13}}\,s s^{-1}$.
The secular decrease is opposite to the expected increase in period bouncers. This together with a L2-L5 type donor detected by Harrison (2016) suggests that WZ Sge is a pre-bounce system. To study its evolution further, we find that the observed $\dot{P}$ is about $1.53$ times larger than by pure GR driving decrease.
We investigate whether WZ Sge is indeed a system still evolving toward its $P_{min}$ by studying the $M_2-P_{orb}$ relation. We constructed the evolution track of WZ Sge, and compare with
the standard model and the revised model track from Knigge et al. (2011). The location of WZ Sge in Fig. 5 would match to $M_2=0.078M_{\odot}$, consistent with Steeghs et al. (2007).
Its evolutionary track predicts a $P_{min}\simeq77.98 (\pm0.90)$ min. Supposing the orbital period decreases at the present rate, then WZ Sge will become a bounce-back CV after $\sim25.3$ Myr. Using the period evolution equation, the mass transfer rate is derived as $\dot{M}_{2}\simeq4.04(\pm0.10)\times10^{-11}M_{\odot}yr^{-1}$, which is compatible with the conclusion of Zharikov et al. (2013).

WZ Sge-type stars were regarded as the dominant CVs at $P_{min}$ (Zharikov et al. 2013; Zharikov 2014). Indeed, these systems have ultrashort orbital periods of $\sim80$ min and undergo rare super-outbursts but normal outbursts are absent. This implies that there is an extremely low viscosity parameter ($\alpha\sim0.01-0.001$) and a very low mass transfer rate (a few$\times10^{-11}M_{\odot}yr^{-1}$) (Smak 1993; Osaki 1994). Already some authors study the properties of accretion disks of the bouncer candidates (e.g. Zharikov et al. 2013; Zharikov 2014). In addition, a few eclipsing CVs with sub-stellar donors were detected (e.g. Littlefair et al. 2006, 2008; Savoury 2011; McAllister et al. 2015). In fact, their eclipsing properties offer important clues concerning the long-term evolution of orbital periods and the identification of period bouncers. This paper provides a good example but much work remains to be done.

\section{Acknowledgments}
This work is supported by the Chinese Natural Science Foundation (Grant No. 11325315, 11611530685, 11573063 and 11133007), the Strategic Priority Research Program ''The Emergence of Cosmological Structure'' of the Chinese Academy of Sciences (Grant No. XDB09010202) and the Science Foundation of Yunnan Province (Grant No. 2012HC011).
This study is also supported by the Russian Foundation for Basic Research (project No. 17-52-53200). New CCD photometric observations of WZ Sge were obtained with the 2.4-m and Sino-Thai 70-cm telescopes in Yunnan Observatories and the 85-cm telescope at Xinglong Observatory in China. Finally, We are very grateful to the anonymous referee for an insightful report that has significantly improved the paper.

\end{document}